\documentclass{PoS}
\usepackage{graphicx}
\usepackage{xspace}

\newcommand{\pp}         {pp\xspace}

\newcommand{\pbpb}       {Pb--Pb\xspace}

\newcommand{\sqrts}      {\ensuremath{\sqrt{s}}\xspace}
\newcommand{\sqrtsnn}    {\ensuremath{\sqrt{s_\mathrm{NN}}}\xspace}
\newcommand{\gevc}       {~GeV/\ensuremath{c}\xspace}

\newcommand{\pip}        {\ensuremath{\pi^+}\xspace}

\newcommand{\pt}         {\ensuremath{p_T}\xspace}

\newcommand{\dndeta}     {\ensuremath{\mathrm{d}N_\mathrm{ch}/\mathrm{d}\eta}\xspace}
\newcommand{\meandndeta} {\ensuremath{\langle\dndeta\rangle}\xspace}

\newcommand{\dens}       {charged-particle pseudorapidity density\xspace}
\newcommand{\npart}      {\ensuremath{N_\mathrm{part}}\xspace}
\newcommand{\ncoll}      {\ensuremath{N_\mathrm{coll}}\xspace}

\title{ALICE soft physics summary}
\ShortTitle{ALICE soft physics summary}

\author{Dariusz Mi\'{s}kowiec for the ALICE Collaboration\\
        GSI Darmstadt and CERN Geneva\\
        E-mail: \email{d.miskowiec@gsi.de}}

\abstract{
Within the first two years of the LHC operation ALICE addressed the major 
soft physics observables in \pbpb and \pp collisions. 
In this contribution we present a selection of these results, with the 
emphasis on the bulk particle production and on particle correlations. 
The latter subject is discussed in detail in several dedicated ALICE talks in 
the same workshop; the reader is referred to the corresponding contributions. 
}
\FullConference{The Seventh Workshop on Particle Correlations and Femtoscopy\\
		 September 20 - 24 2011\\
		 University of Tokyo, Japan}

\begin{document}
\section{Introduction}
\label{intro}
\begin{figure}[b]
\hspace*{-2mm}\rotatebox{270}{\scalebox{0.52}{\includegraphics{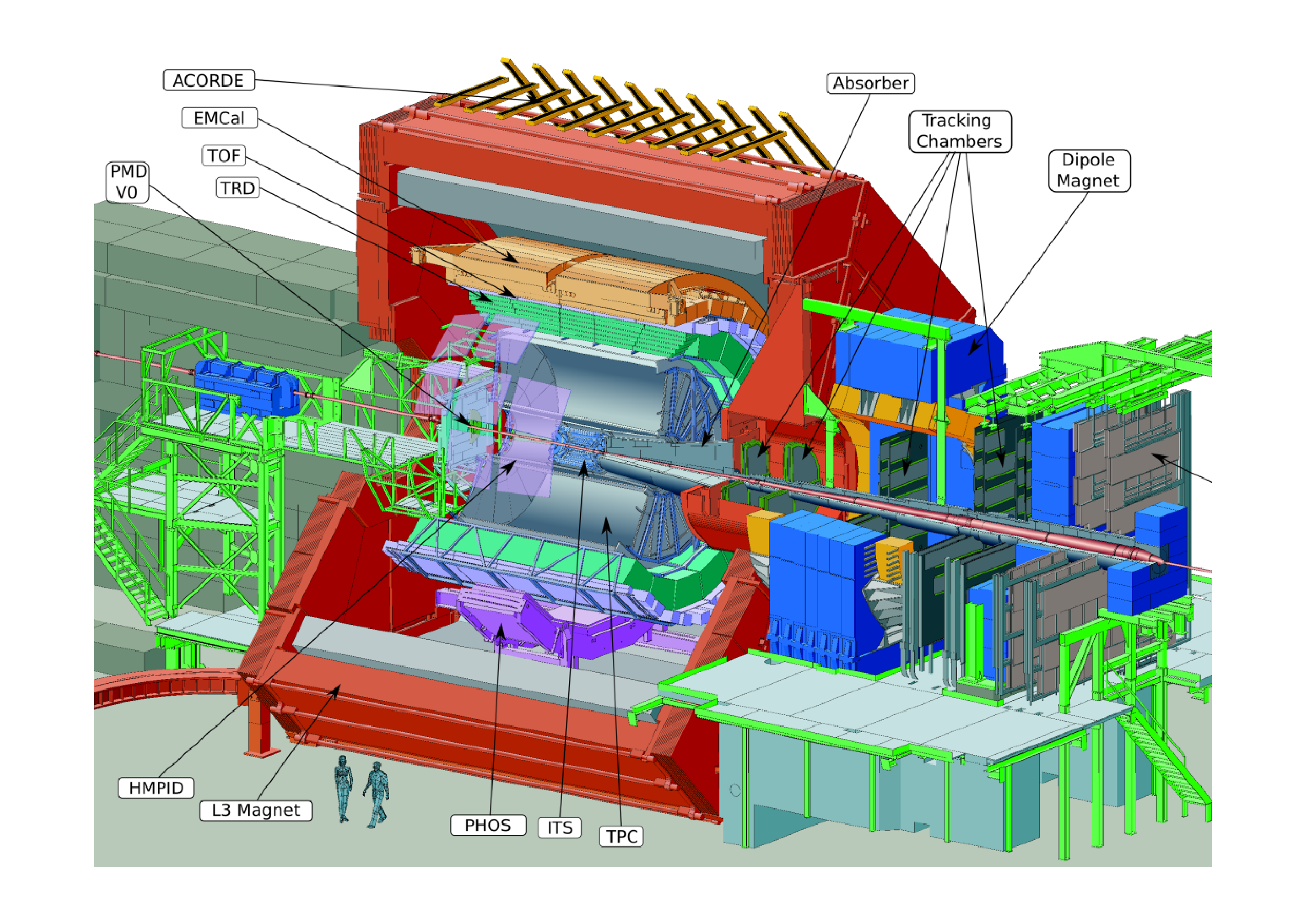}}}
\caption{The ALICE experiment at the CERN LHC. The central-barrel detectors 
(ITS, TPC, TRD, TOF), with a pseudorapidity coverage $|\eta|<0.9$, address 
the particle production at midrapidity. The centrality is determined from 
the charged-particle multiplicity at $1.7<|\eta|<5.1$.}
\label{fig:setup}
\end{figure}

ALICE (A Large Ion Collider Experiment) is an experiment at the Large Hadron 
Collider dedicated to studies of QCD matter created in energetic collisions 
between lead nuclei~\cite{Carminati:2004fp}. 
QCD predicts that at energy densities above 1 GeV/fm$^3$ a state of deconfined 
quarks and gluons occurs, possibly accompanied by chiral symmetry restoration 
in which quarks assume their current masses. 
Assessing the properties of the created matter requires sound understanding of 
the underlying collision dynamics which can be best studied via the bulk 
particle production observables. 
ALICE has a high granularity, a low transverse momentum threshold 
$\pt^\mathrm{min} \approx0.15$~\gevc, and a good hadron identification up to 
several\gevc and is thus perfectly suitable for addressing soft physics 
observables in heavy-ion collisions. 
The setup is shown in Fig.~\ref{fig:setup}. 
The central-barrel detectors, the Inner Tracking System (ITS), the Time Projection 
Chamber (TPC), the Transition Radiation Detector (TRD), and the Time Of Flight (TOF), 
cover the full azimuthal angle range at midrapidity $|\eta|<0.9$. 
The calorimeters EMCal and PHOS and the particle identification detector HMPID 
have partial coverage at midrapidity. 
The V0 detector measures charged-particle multiplicity at $-3.7<\eta<-1.7$ and 
$2.8<\eta<5.1$ and is mainly used for triggering and centrality determination. 
Several other detector systems exist but are not relevant to the results 
discussed in these proceedings. 
The collision systems and energies measured by ALICE are summarized in Table~\ref{tab:data}. 
\begin{table}[h]
\caption{Data sets collected by ALICE in 2009--2012 (excluding rare triggers). }
\label{tab:data}
\begin{tabular*}{\textwidth}{@{}c@{\extracolsep{\fill}}cccc}
\hline
year & system & \sqrtsnn (TeV) & trigger & \# events/$10^6$\\
\hline
2009 & \pp   & 0.9  & min bias & 0.3 \\
2009 & \pp   & 2.36 & min bias & 0.04 \\
\hline
2010 & \pp   & 0.9  & min bias & 8 \\
2010 & \pp   & 7.0  & min bias & 800 \\
     &       &      & high multiplicity & 50 \\
     &       &      & forward muons & 50 \\
2010 & \pbpb & 2.76 & min bias & 30 \\
2010 & \pp   & 2.76 & min bias & 70 \\
\hline
2011 & \pp   & 7.0  & min bias & $\sim$$10^3$ \\
2011 & \pbpb & 2.76 & central & $\sim$30 \\
     &       &      & semicentral & $\sim$30 \\
     &       &      & min bias & $\sim$10 \\
     &       &      & forward muons & $\sim$20 \\
\hline
\end{tabular*}
\end{table}

\section{Charged-particle production}
\label{charged}
The bulk particle production is a basic indicator of entropy during a nuclear 
collision. 
The \dens\ in central (0-5\%) \pbpb\ collisions at \sqrtsnn=2.76~TeV, 
normalized to the number of participant pairs, is $\dndeta/(0.5\langle\npart\rangle)=8.3\pm0.4$, 
about two times higher than in \pp collisions at the same energy, 
\begin{figure}[b]
\hspace{2.5cm}\scalebox{0.48}{\includegraphics[clip]{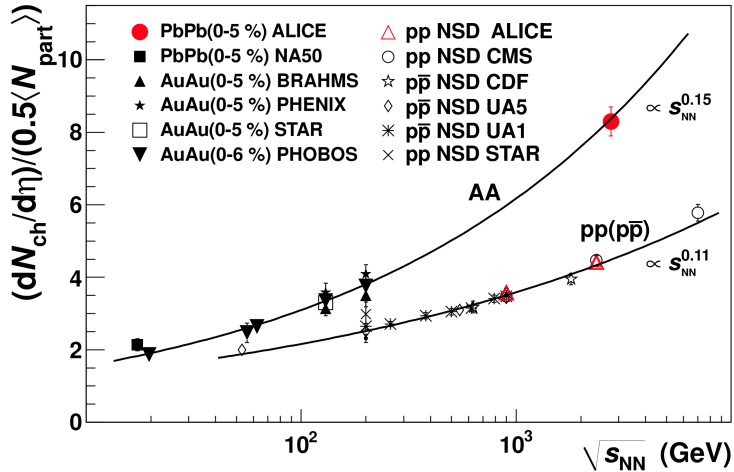}}
\caption{Charged-particle yields at midrapidity from \pp and heavy-ion collisions~\cite{Aamodt:2010pb}.}
\label{fig:charged}
\end{figure}
and also about twice as high as in the gold--gold collisions at RHIC (Fig.~\ref{fig:charged}).  
Both in \pp and \pbpb\ the \dens\ is a power-law function of \sqrts. 
The exponent is higher in nuclear collisions. 
The observed \dens exceeds the predictions of most models with the initial state gluon saturation. 
The bulk particle production was subject of the first ALICE publication from 
the lead campaign at the LHC~\cite{Aamodt:2010pb}. 

\begin{figure}[h]
\hspace{2cm}\includegraphics[width=10cm,clip]{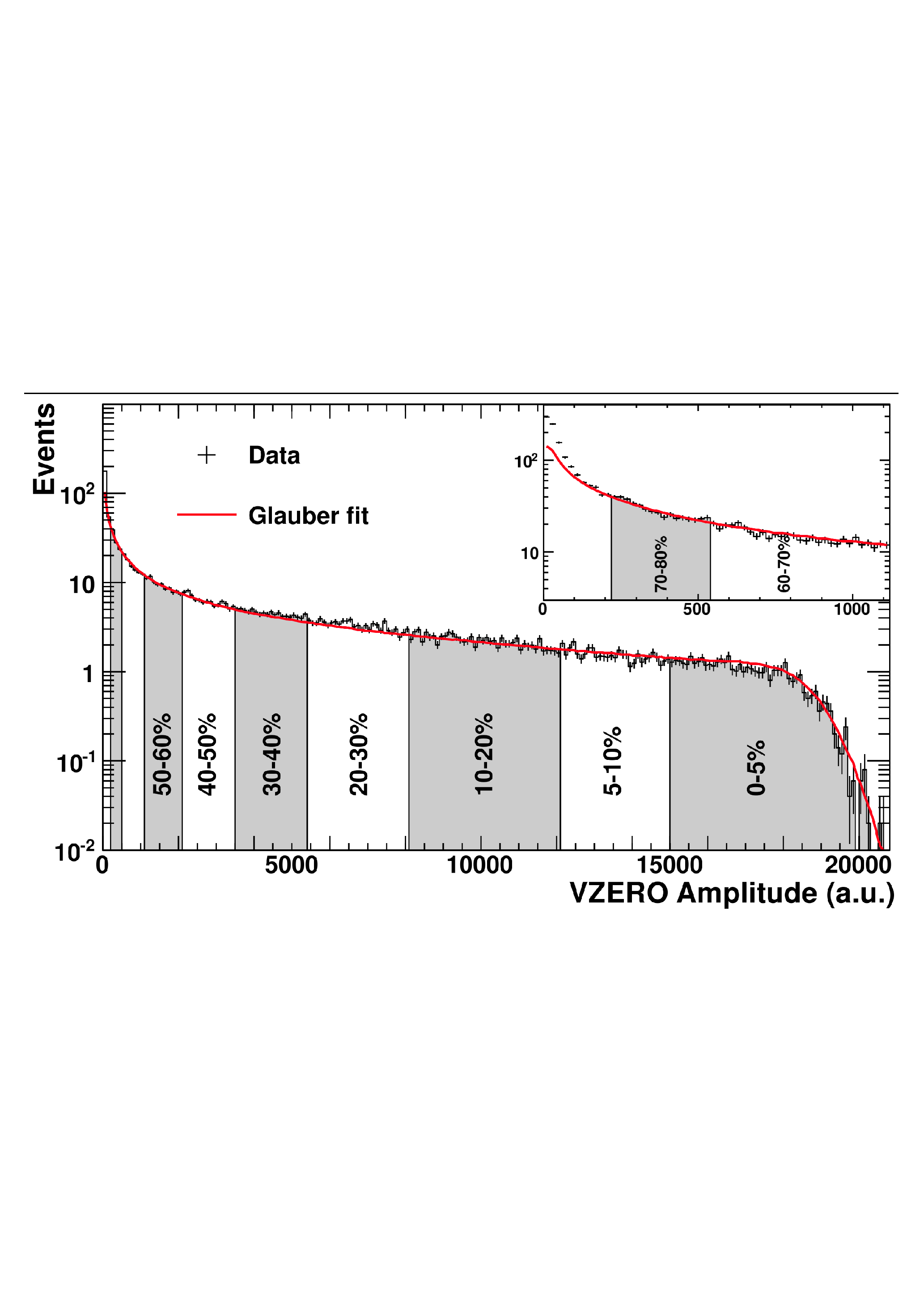}
\caption{Determination of the centrality using the charged-particle multiplicity 
at $1.7<|\eta|<5.1$~\cite{Aamodt:2010cz}.}
\label{fig:centrality}
\end{figure}

More information is contained in the centrality dependence of the \dens. 
The centrality of the collision events, in terms of the fraction of 
the geometric cross section, was derived from the charged-particle multiplicity 
seen in the V0 detector at $1.7<|\eta|<5.1$. 
For this, the multiplicity distribution (Fig.~\ref{fig:centrality}) 
was fitted by a simple model assuming $f\npart+(1-f)\ncoll$ particle sources, 
each source producing particles following a negative binomial 
distribution (red line). 
The number of participants and the number of binary collisions, \npart and \ncoll, were 
calculated for each impact parameter using Glauber model. 
The centrality resolution was better than 1\%. For the details of this 
analysis see Ref.~\cite{Aamodt:2010cz}. 

\begin{figure}[b]
\hspace{3cm}\scalebox{0.45}{\includegraphics[clip]{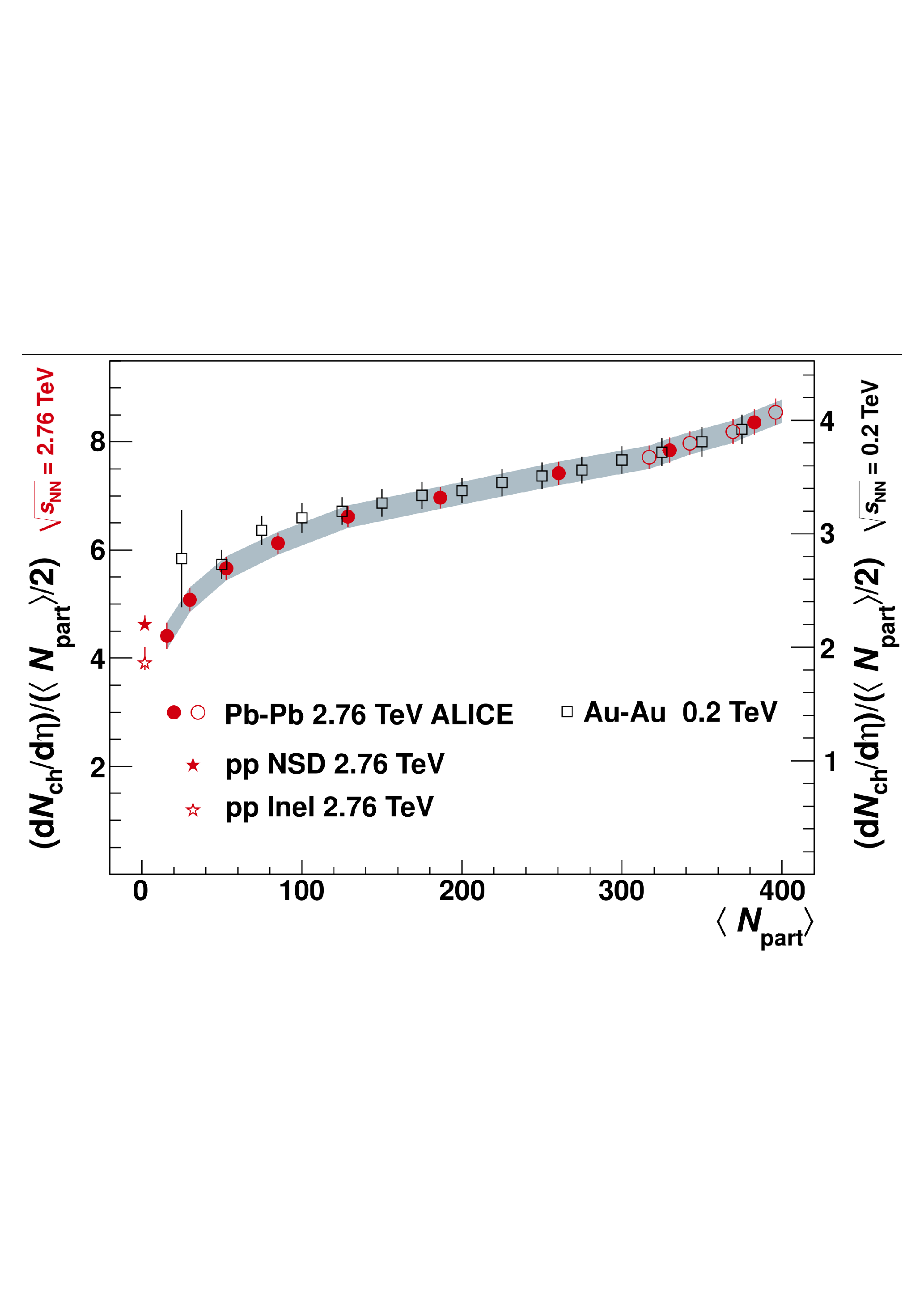}}\\
\vspace{-0.8cm}
\caption{Charged-particle production as a function of centrality~\cite{Aamodt:2010cz}. 
The LHC measurement coincides with the scaled RHIC data.}
\label{fig:npart}
\end{figure}

The centrality dependence of the normalized \dens 
turns out to coincide in shape with the one measured at RHIC (Fig.~\ref{fig:npart}). 
This is against the expectation that an increased contribution of 
hard processes should lead to a steeper centrality dependence at the LHC. 
The weak centrality dependence observed is presumably related to the nuclear 
shadowing and is, in fact, reasonably well reproduced by models which take this 
phenomenon into account~\cite{Muller:2012zq}. 

\section{Identified hadrons}
\label{ident}

\begin{figure}[h]
\hspace{3.5cm}\scalebox{0.36}{\includegraphics[clip]{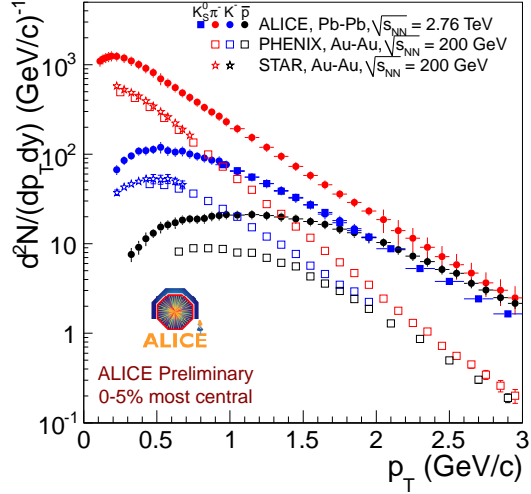}}
\caption{Transverse momentum spectra of identified hadrons from ALICE.}
\label{fig:spectra}
\end{figure}

A better handle on the reaction dynamics can be obtained using 
identified hadrons. 
Transverse momentum spectra of pions, kaons, and protons from \pbpb collisions 
are shown in Fig.~\ref{fig:spectra}. 
The spectra at the LHC are harder than at RHIC~\cite{Floris:2011ru}. 
A blast wave fit allows one to associate this fact with a 10\% increase 
of the average transverse flow velocity. 

\begin{figure}[b]
\hspace{0.8cm}\scalebox{0.33}{\includegraphics[clip]{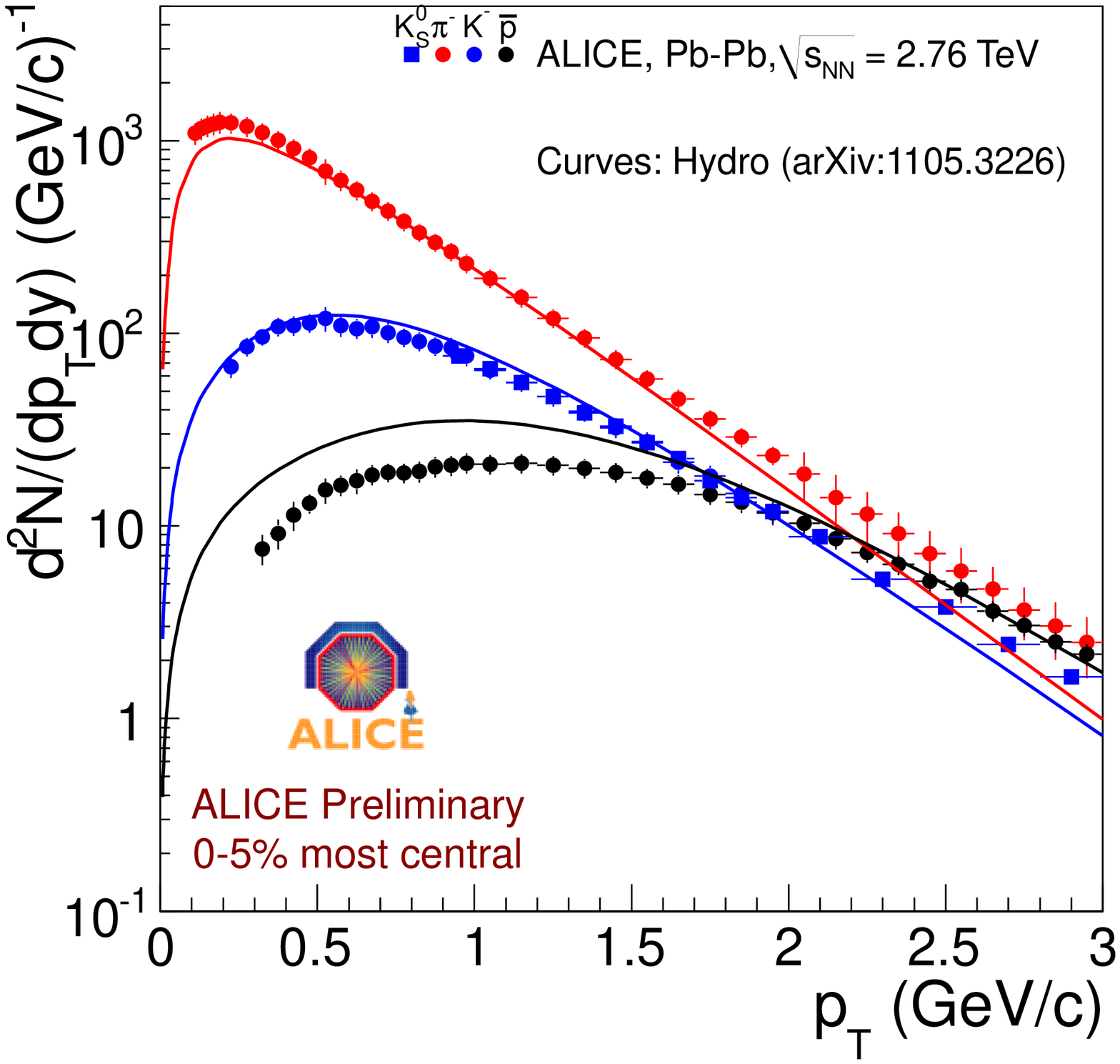}}
\hspace{1cm}\scalebox{0.33}{\includegraphics[clip]{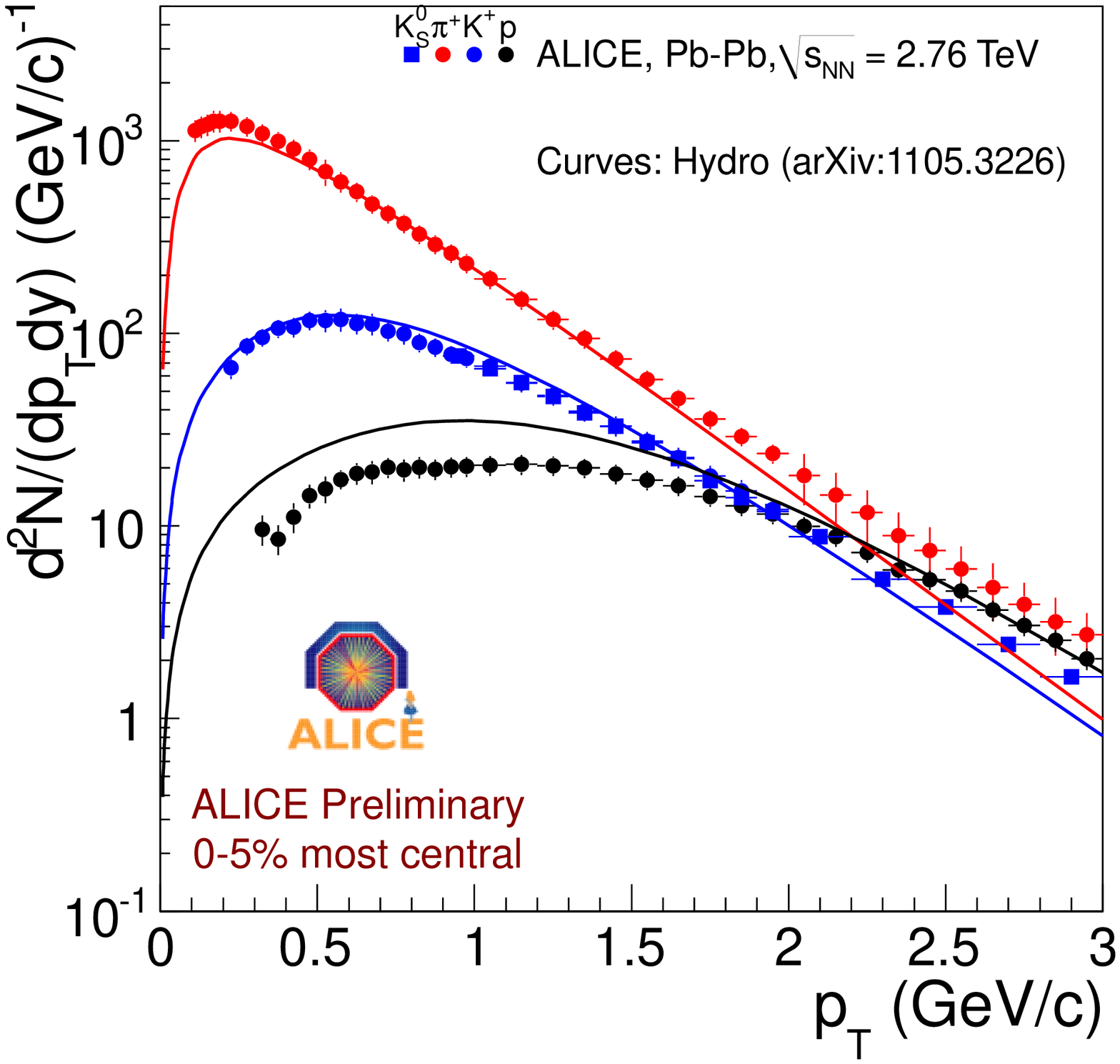}}
\caption{Transverse momentum spectra of identified hadrons compared to 
hydro predictions.}
\label{fig:spectrahydro}
\end{figure}

Hydrodynamic predictions~\cite{Shen:2011eg} underestimate the high-\pt 
part (Fig.~\ref{fig:spectrahydro}). 
They also overestimate proton yield which might indicate a too high 
chemical freeze-out temperature $T_{\rm ch}$. 
A similar discrepancy is present in the thermal model. There, however, 
a lower $T_{\rm ch}$ is excluded by the $\Xi$ and $\Omega$ yields~\cite{kalweit}. 
The details of the identified hadron analysis are given in Ref.~\cite{Floris:2011ru}. 

Another exciting prospect is to look for production of light nuclei and 
hypernuclei. Combining the specific energy loss in the TPC with the time 
of flight from TOF, ALICE was able to identify four anti-alpha particles 
in the 2010 \pbpb data set. An order of magnitude more abundant was the 
antihypertriton $^3_{\overline\Lambda}\overline{\rm H}$ detected via its 
decay into $^3\overline{\rm He}$ and \pip. For antihelium $^3\overline{\rm He}$, 
a transverse momentum spectrum was measured in the range of 1-8\gevc. 
The search for such composed objects with ALICE is discussed in Refs.~\cite{Kalweit:2011if,Sharma:2011yf,lea}. 

\section{Femtoscopy}
\label{femto}
\begin{figure}[h]
\hspace{2.5cm}\scalebox{0.65}{\includegraphics[clip]{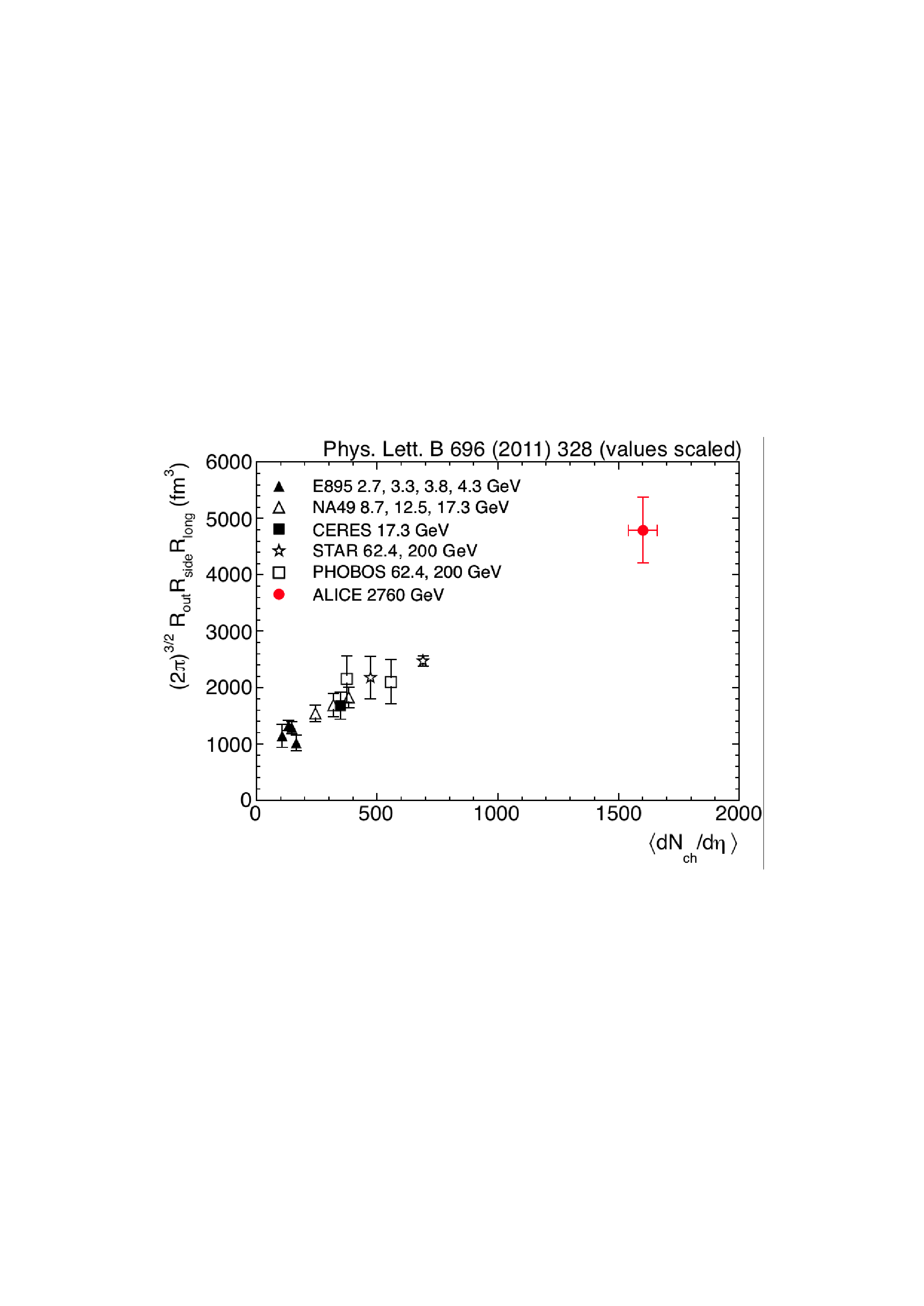}}
\caption{Homogeneity volume in central gold and lead collisions at different 
energies~\cite{Aamodt:2011mr}.}
\label{fig:hbt1}
\end{figure}

An increase of the spatial extension of the particle source from RHIC to LHC energies 
was declared a decisive test of whether the concept of ``matter'' is at all 
applicable to the system created in energetic nuclear collisions~\cite{Satz:2011wf}. 
Spatial extension of the source of pions with a fixed momentum vector 
(homogeneity length, or Hanbury Brown--Twiss (HBT) radius) is accessible via the 
Bose-Einstein correlations aka HBT analysis. 
The pion homogeneity volume measured by ALICE in most central 5\% \pbpb collisions 
at \sqrtsnn=2.76~TeV~\cite{Aamodt:2011mr} is shown in Fig.~\ref{fig:hbt1}. 
The volume is about two times larger than the analogous volume at RHIC and 
scales linearly with multiplicity. 

\begin{figure}[h]
\hspace{3cm}\scalebox{0.44}{\includegraphics[clip]{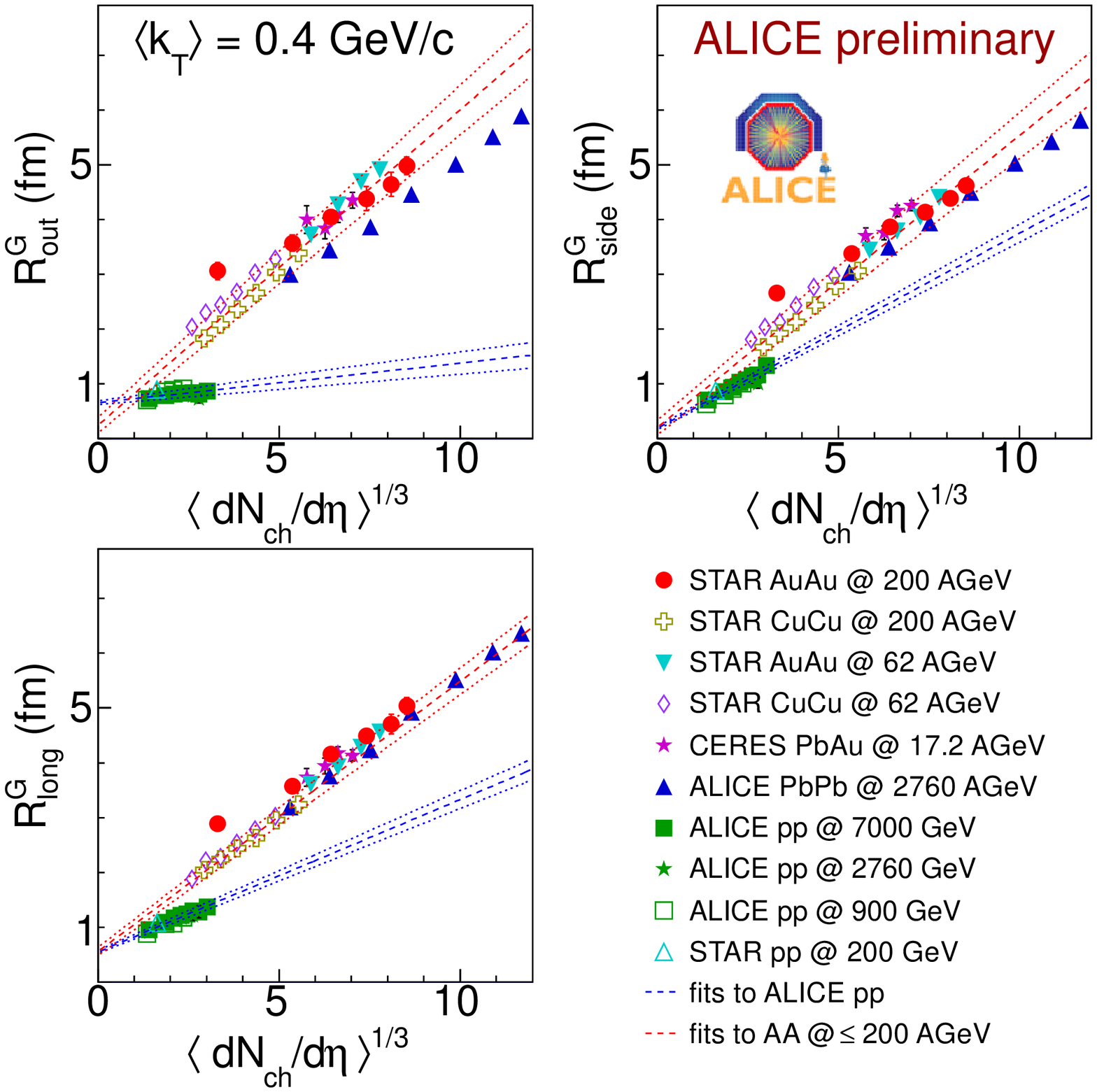}}
\caption{Scaling of HBT radii with the \dens in pp and AA collisions.}
\label{fig:hbt2}
\end{figure}

While the homogeneity volume of central heavy-ion collisions observed at various 
collision energies nicely scales with the \dens, including other centralities and, 
in particular, other collision systems leads to a violation of this scaling. 
This is particularly clear when comparing the volumes measured at the LHC 
in \pp and \pbpb at the same \dens (Fig.~\ref{fig:hbt2})~\cite{Aamodt:2011kd,wpcf-kisiel}. 
The three HBT radii always scale linearly with the cube root of \meandndeta but the 
slope of the scaling is different for protons and heavy-ions. 
This indicates that the HBT radii are not driven exclusively by the final multiplicity 
but are also sensitive to the initial geometry of the collision. 

\begin{figure}[h]
\hspace{4.5cm}\scalebox{0.29}{\includegraphics[clip]{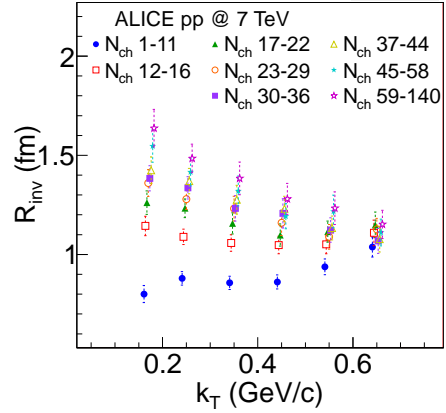}}
\caption{Evolution of the transverse momentum dependence of HBT radii with multiplicity in 
pp collisions~\cite{Aamodt:2011kd}.}
\label{fig:hbt3}
\end{figure}

The collective outward motion of particles at freeze-out is the commonly accepted 
explanation for the fact that the HBT radii in heavy-ion collisions decrease with 
increasing transverse momentum. 
The observation of a similar \pt dependence in \pp collisions at RHIC suggested that 
either this explanation has to be revised, or collective flow exists also in \pp collisions. 
Some clarification on the subject and support to the latter statement comes from the 
observation that the \pt dependence in \pp collisions develops with increasing 
multiplicity~\cite{Khachatryan:2011hi,Aamodt:2011kd}. This is shown in Fig.~\ref{fig:hbt3} 
and discussed in detail in Refs.~\cite{Aamodt:2011kd,wpcf-kisiel,wpcf-humanic}. 

\section{Elliptic flow}
\label{flow}
\begin{figure}[h]
\hspace{2.7cm}\scalebox{0.5}{\includegraphics[clip]{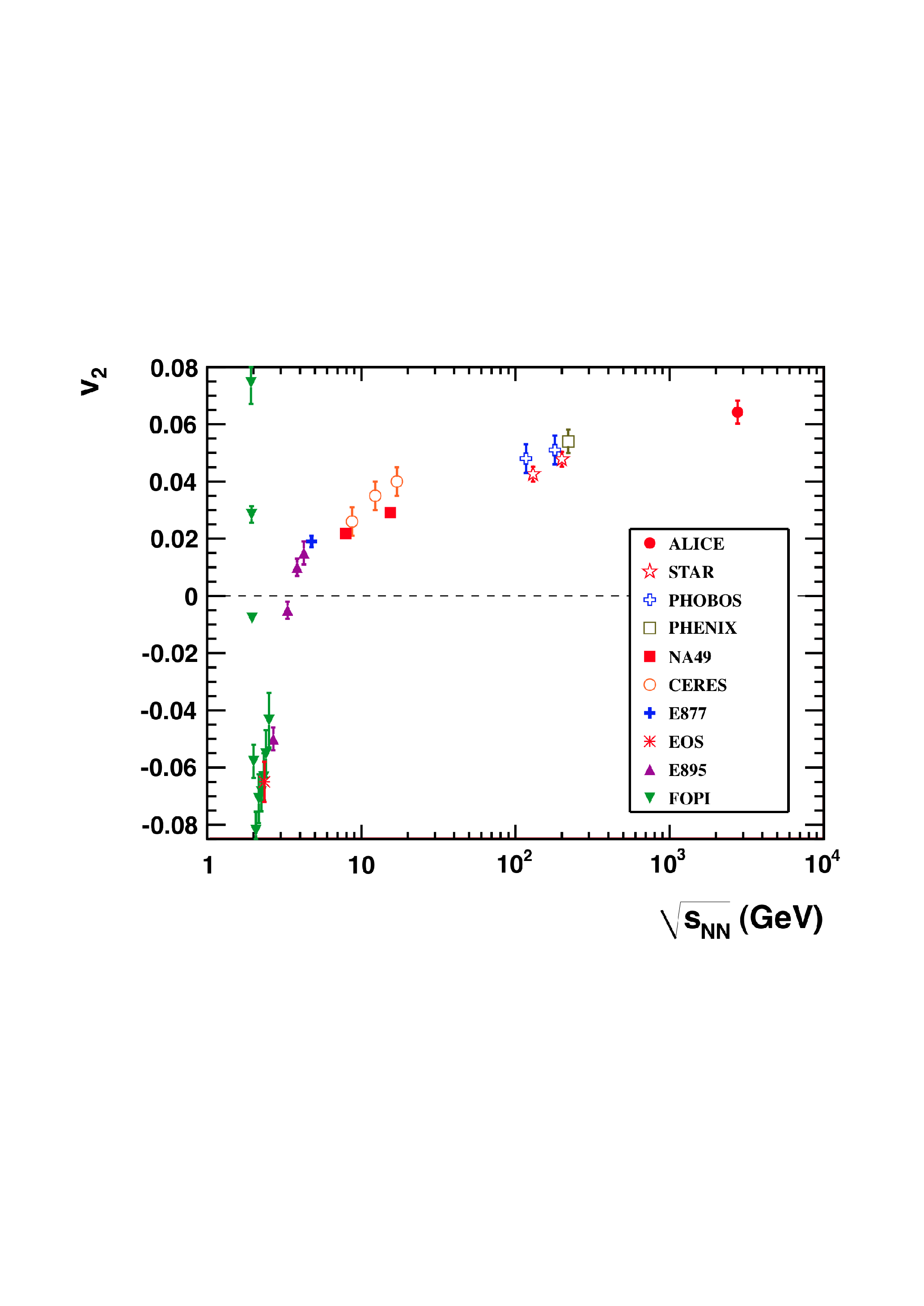}}
\caption{Collision energy dependence of the elliptic flow~\cite{Aamodt:2010pa}. 
The ALICE measurement matches the trend established at lower energies.}
\label{fig:v2}
\end{figure}

After the RHIC experiments found that the QCD matter created in heavy-ion 
collisions behaved like a low-viscosity fluid, a natural question arose 
whether this behavior would continue at higher collision energies, or whether 
the system would get closer to a non-interacting gas of quarks and gluons. 
This question was answered by the ALICE measurement of the elliptic flow coefficient 
$v_2$ which quantifies the azimuthal anisotropy of the particle emission in non-central 
collisions and which, therefore, is sensitive to the early stage of the 
collision~\cite{Aamodt:2010pa}. As is shown in Fig.~\ref{fig:v2}, 
the elliptic flow at the LHC turned out to be higher than at RHIC and to 
follow the trend observed at lower energies. 
The $v_2$ increase at the LHC mostly comes from the increase in $\langle\pt\rangle$,  
as the shape of the $v_2(\pt)$ dependence remains unchanged within the \sqrtsnn 
range from 40 to 2760~GeV. 

\begin{figure}[b]
\hspace{2.5cm}\scalebox{0.47}{\includegraphics[clip]{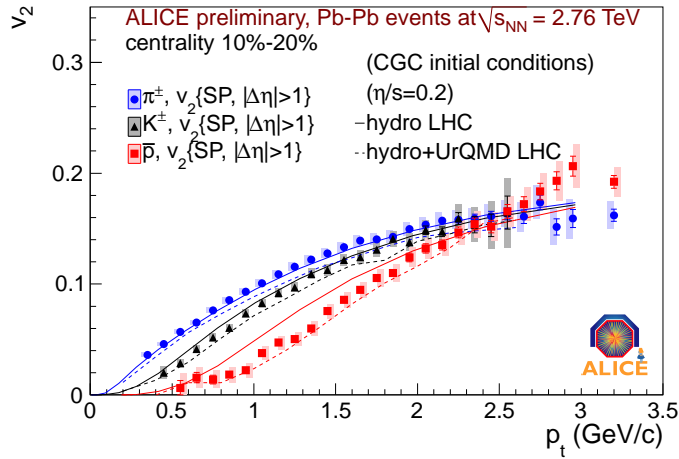}}
\caption{Transverse momentum dependence of the elliptic flow of pions, kaons, 
and protons. The splitting is presumably caused by transverse radial flow.}
\label{fig:v2pthydro}
\end{figure}

The elliptic flow coefficient of identified hadrons shows a splitting characteristic 
to the presence of transverse radial flow (Fig.~\ref{fig:v2pthydro})~\cite{noferini,wpcf-ivan}. 
The shape of the \pt dependence and the presence of the splitting are fairly well 
reproduced by hydrodynamics~\cite{Heinz:2011kt}. The calculation, however, 
underpredicts the elliptic flow of protons. This can be cured by adding hadronic 
rescattering~\cite{Heinz:2011kt}. 

An important role at the LHC is played by the triangular flow. Its coefficient 
$v_3$ only weakly depends on centrality and its direction is not related to 
the orientation of the event plane which points to the initial energy density 
fluctuations as its origin. 
There are indications that the ``Mach cone''-like azimuthal emission pattern is 
actually just a superposition of the elliptic and triangular collective 
flow~\cite{ALICE:2011ab,wpcf-sano,wpcf-ivan}. 

\section{Fluctuations}
\label{fluct}

\begin{figure}[h]
\hspace{0cm}\rotatebox{270}{\scalebox{0.42}{\includegraphics[clip]{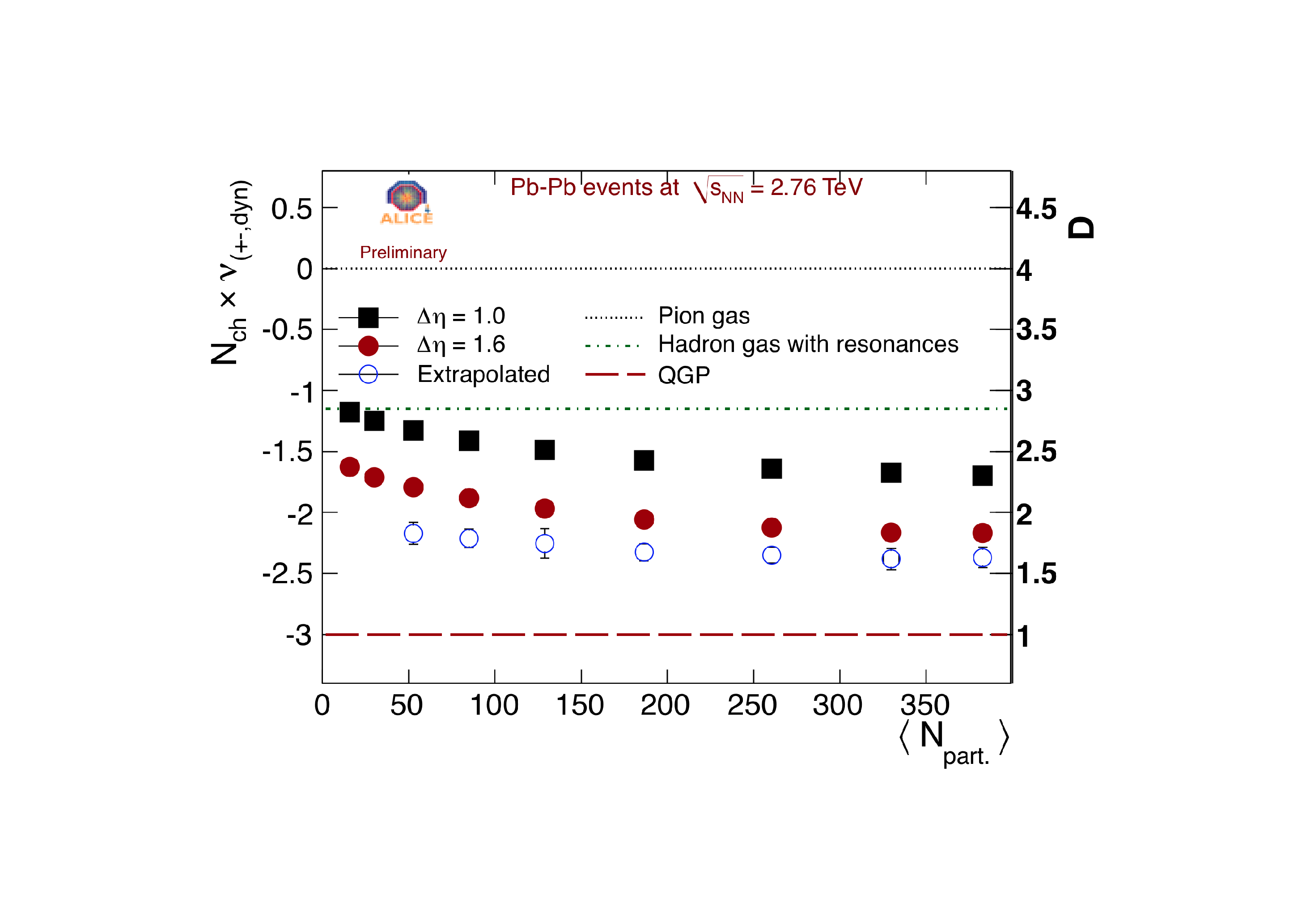}}}

\vspace*{-5.9cm}\hspace{9cm}\scalebox{0.395}{\includegraphics[clip]{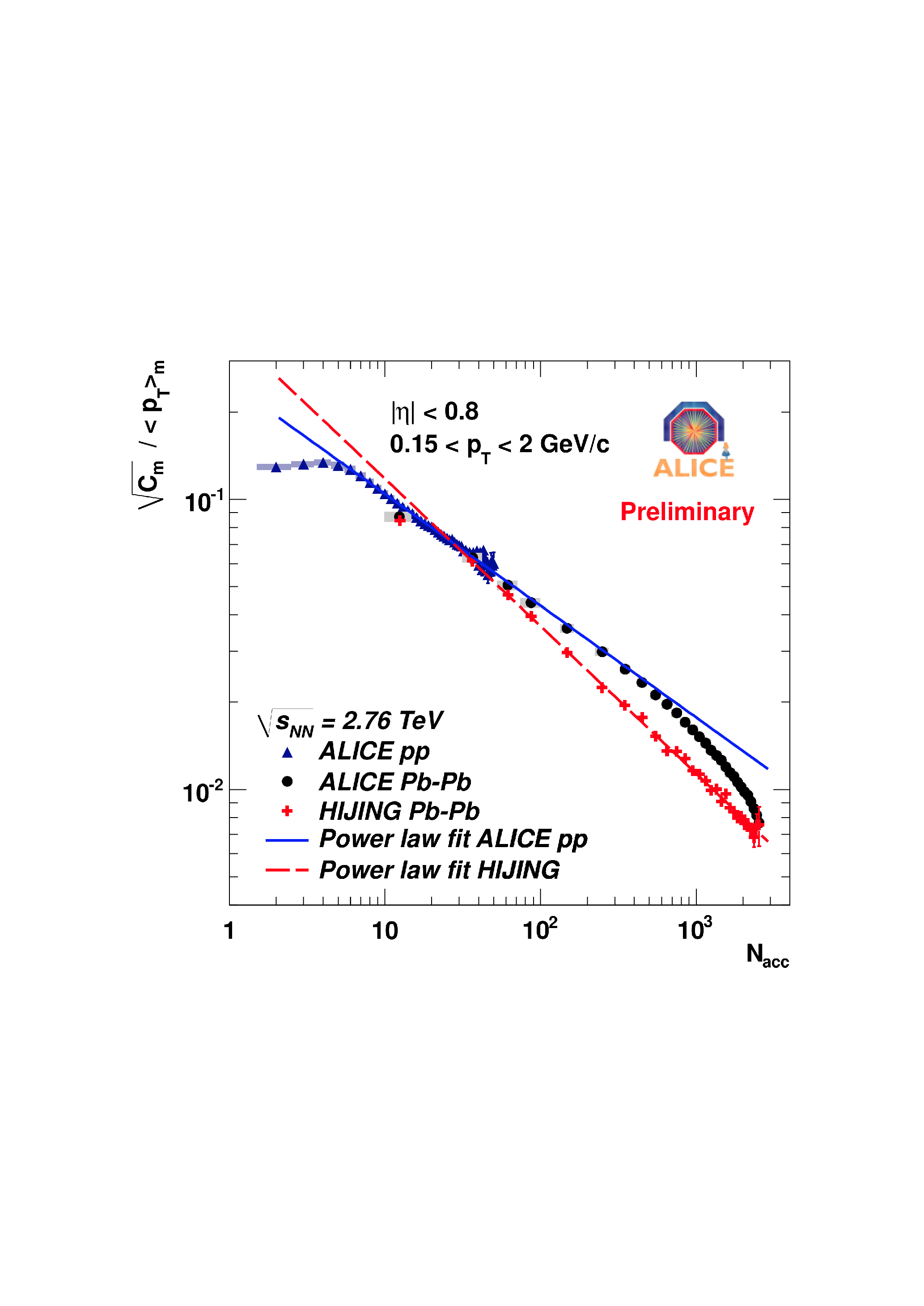}}
\caption{Multiplicity dependence of charge (left) and \pt (right) fluctuations 
measured by ALICE.}
\label{fig:fluct}
\end{figure}

The magnitude of charge fluctuations should reflect the number of charge 
carriers, and thus be different for QGP and hadron gas. 
It is puzzling that the experimental results so far were closer to the hadron gas 
calculation than to QGP (although the distinction is somewhat obscured by resonances 
that introduce anticorrelations between positive and negative charges). 
In an ALICE analysis of \pbpb at \sqrtsnn=2.76~TeV the net-charge fluctuations, 
expressed in terms of 
$\nu_{(+,-,{\rm dyn})} = \left\langle\left(\frac{N_+}{\langle N_+ \rangle}-\frac{N_-}{\langle N_- \rangle}\right)^2\right\rangle 
- \frac{1}{\langle N_+ \rangle} - \frac{1}{\langle N_- \rangle} $
and corrected for the finite acceptance, approach the QGP value in most central 
collisions as shown in the left panel of Fig.~\ref{fig:fluct} 
(see Ref.~\cite{wpcf-jena} for details). 

Enhanced transverse momentum fluctuations may signal vicinity to the critical 
point of the QCD phase transition. 
The ALICE measurement of the \pt fluctuations,
expressed via the mean covariance between transverse momenta of track pairs $C_m$, 
is shown for \pbpb and \pp collisions in the right hand panel of 
Fig.~\ref{fig:fluct}. 
Unlike the HBT radii discussed above, the relative \pt fluctuations 
fall on a universal curve when plotted against multiplicity. 
The dependence is of the power-law type, except for deviations at very soft \pp 
and at central \pbpb collisions. 
The latter is not reproduced by the HIJING event generator. 
The details of this analysis are given in Ref.~\cite{Stefan:2011es}. 

The transient magnetic field of two ions colliding at a finite impact 
parameter may lead to charge-dependent angular correlations. 
The STAR experiment at RHIC reported such correlations developing 
when going from central to peripheral collisions of gold nuclei~\cite{:2009uh}. 
The analogous measurement in \pbpb performed by ALICE agrees within errors 
with the STAR result~\cite{wpcf-selyuzhenkov}. 

\section{Summary}
\label{summary}

During the first campaign of \pbpb collisions at the LHC, ALICE addressed the most 
important soft physics observables. The lead collision studies were augmented 
by \pp measurements at several energies. 
New insights into the reaction dynamics include an alternative interpretation of 
the angular emission pattern (flow rather than the ``Mach cone''), the \pt 
dependence of HBT radii developing with 
multiplicity in \pp collisions, and the proton puzzle (lower than expected yield 
and elliptic flow). 

As the c.m.s. energy increase from RHIC to LHC is unprecedentedly large 
it is interesting to verify whether the energy dependence trends found at lower 
energies are still valid at the LHC. 
The results discussed in this paper fall in three categories. 
First, about two times larger than at RHIC are the particle yield 
and the homogeneity volume. Second, an increase by 10-30\% is 
observed in the transverse flow, mean transverse momentum, integrated 
elliptic flow, and the mass splitting of $v_2$. 
(The latter three may be a consequence of the first.) 
Finally, unchanged with respect to RHIC remained the centrality dependencies 
of the particle yield and of $v_2$, the multiplicity dependencies of pion HBT 
radii and of particle ratios, the \pt dependence of $v_2$, and the charge and 
\pt fluctuations. 

The author thanks the organizers for the pleasant and inspiring atmosphere 
during the meeting. 



\end{document}